# Resonance results in 7 TeV pp collisions with the ALICE detector at the LHC

Massimo Venaruzzo for the ALICE Collaboration[a,*]

[a]*Università degli Studi di Trieste and INFN Sezione di Triese, Via A. Valerio 2, 34127 Trieste, Italy.*

**Abstract**

Short lived hadronic resonances are a very important tool for the study of the dynamics of the matter produced in heavy-ion collisions since they should be sensitive to the medium properties such as temperature, density and expansion velocity. In particular they are sensitive to the time span between chemical and kinetic freeze-out of the hadronic phase of the fireball. The study of resonances in 7 TeV pp collisions is useful both in order to constrain QCD inspired models and to form a baseline for the production in heavy-ion collisions. The resonances $K^{*0}(892)$, $\phi(1020)$, $\Sigma(1385)^{\pm}$, $\Lambda(1520)$, $\Xi(1530)$ are reconstructed from their hadronic decay using data collected by the ALICE detector in pp collisions at 7 TeV. Their yields and $p_T$ spectra are compared with Monte Carlo models such as PHOJET and different PYTHIA tunes.

## 1. Introduction

Resonance production is an important issue both in pp and in heavy-ion collisions. In pp collisions, it contributes to the understanding of the underlying event and the hadron production [1] and it provides a reference for tuning QCD-inspired event generators. In heavy-ion collisions, due to their short lifetime (few fm/$c$), resonances can both decay and be regenerated by final state interactions inside the hot and dense matter and therefore are sensitive to its dynamical evolution [2–4]. The hadronic resonances $K^{*0}(892)$, $\phi(1020)$, $\Sigma(1385)^{\pm}$, $\Lambda(1520)$ and $\Xi(1530)$ have been measured by the ALICE experiment in pp collisions at an energy in the center of mass of $\sqrt{s} = 7$ TeV. Data analysis was carried out at midrapidity ( $|y| <$ 0.5 for $K^{*0}(892)$ and $\phi(1020)$, $|y| < 0.8$ for $\Sigma(1385)^{\pm}$, $\Lambda(1520)$ and $\Xi(1530)$ ) on a sample of minimum-bias events collected by ALICE during the CERN LHC run in 2010. In section 2 a brief introduction of the detector characteristics which were used for this analysis is presented. The raw yield extraction procedure is described in section 3. In section 4 the corrected transverse momentum spectra for $\phi(1020)$ and $\Sigma(1385)^{\pm}$ are compared to two QCD based event generators, PHOJET [5] and PYTHIA [6]. Finally, in section 6 the ratios $\phi/K^-$, $\phi/\pi^-$, $(\Omega + \bar{\Omega})/\phi$ are calculated with the yield of $\pi$, K and $\Omega + \bar{\Omega}$ previously measured with the ALICE detector [7, 8]. The ratios $\phi/K^-$ and $\phi/\pi^-$ are compared with previous measurements at lower energies. Conclusions are drawn in section 7.

## 2. Experimental setup

The ALICE detector is described in detail in [9, 10]. For the analyses presented here, only the central barrel has been used.

---

*Speaker
 Email address: massimo.venaruzzo@ts.infn.it (Massimo Venaruzzo for the ALICE Collaboration)

The central tracking and PID detectors, covering a pseudorapidity window of $|\eta| < 0.9$, include, from the innermost outwards, the Inner Tracking System (ITS), the Time Projection Chamber (TPC) and the Time of Flight array (TOF). The central detectors are embedded in a 0.5 T solenoidal field. The moderate field, together with a low material budget permits the reconstruction of low $p_T$ tracks. The ITS is composed of six layers of silicon detectors. The TPC [11] provides track reconstruction with up to 159 three-dimensional space points per track in a cylindrical active volume of about 90 m$^3$. In order to improve the global resolution, tracks were accepted only in the range $|y| < 0.8$ (i.e. well within the TPC acceptance) and with $p_T > 0.15$ GeV/$c$. The standard tracking used in this analysis combines the information from the ITS and TPC. It provides very good resolution in the distance of closest approach to the vertex, and hence better separation of primary and secondary particles. The TPC identifies particles via the specific energy loss d$E$/d$x$ with a 5.2% resolution [11]. The TPC d$E$/d$x$ measurement allows pions to be separated from kaons for momenta up to $p\sim0.7$ GeV/$c$, while the proton/antiproton band starts to overlap with the pion/kaon band at $p\sim1$ GeV/$c$. Placed at a radius of 370-399 cm, the TOF measures the time-of-flight of the particles, allowing identication at higher $p_T$. With a total time resolution of about 160 ps, pions can be separated from kaons up to $p\sim1.5$ GeV/$c$ and the two mesons can be distinguished from (anti)protons up to $p\sim2.5$ GeV/$c$.

## 3. Analysis Description

The data analysis is carried out using a sample of minimumbias pp data collected by ALICE during 2010, with a size ranging from 60 to 210 million events, for the different resonances analyzed. The resonances have been identified via their main decay channel: $K^{*0}\rightarrow\pi^{\pm} + K^{\mp}$, $\phi\rightarrow K^+K^-$, $\Sigma^{*\pm}\rightarrow\Lambda + \pi^{\pm}$, $\Xi^{*\pm}\rightarrow\Xi + \pi^{\pm}$ and $\Lambda^*\rightarrow N + K$ Due to their very short lifetime, resonance decay products cannot be distinguished from the particles coming from the primary vertex, and their yield can only be measured by first computing the invariant mass spectrum of all primary candidate tracks and then subtracting the combinatorial background.

*3.1. Signal Extraction*

Different signal extraction methods were implemented, but here, for brevity reasons, just the main will be shown. For the φ(1020), Σ(1385)$^{\pm}$, Λ(1520) and Ξ(1530) the combinatorial background was evaluated using the Event-Mixing technique, while for the K$^{*}$ using the Like-Sign technique. The signal after the backgroung subtraction was then fitted with a Voigtian[1] (for the φ(1020) and Ξ(1530)) and a Breit-Wigner (for the K$^{*0}$, Σ(1385)$^{\pm}$ and Λ(1520)) function plus a polynomial for the residual background (Figures 1, 2, 3, 4, 5). For the φ(1020) and the Ξ(1530) one parameter, namely the Gaussian standard deviation $\sigma$ for the φ(1020) and the Breit-Wigner width Γ for the Ξ(1530), were constrained using the PDG value [12]. In all cases the masses, and the widths are compatible within errors with the corresponding PDG values from [12]. Alternative methods, like the combined fit and the like-sign, have been employed with comparable success for the description of the combinatorial background. After the signal is extracted, the raw yield is evalutated as the integral of the signal function. The raw yields extracted in different $p_T$ bins are corrected for the efficiency and the acceptance and the differential transverse momentum spectra are obtained but here only the φ(1020) and Σ(1385)$^{\pm}$ spectra will be shown.

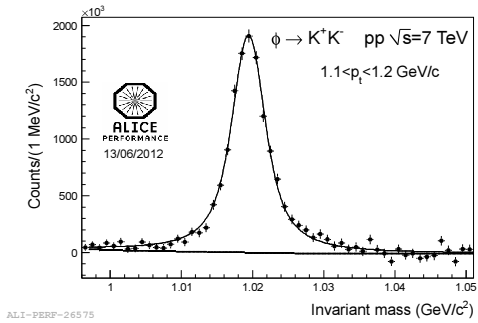

Figure 1: φ(1020) signal after the background subtraction. The fitting function (in black) is the sum of a polynomial and a Voigtian. The background polynomial from the fit is dashed in black.

## 4. Results

The transverse momentum spectrum of the φ is shown in Figure 6[2].

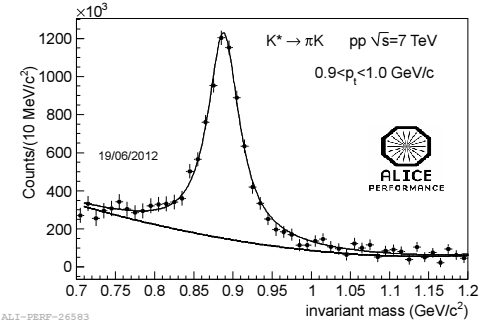

Figure 2: K$^{*}$ signal after the background subtraction. The fitting function (in black) is the sum of a polynomial and a Voigtian. The background polynomial from the fit is dashed in black.

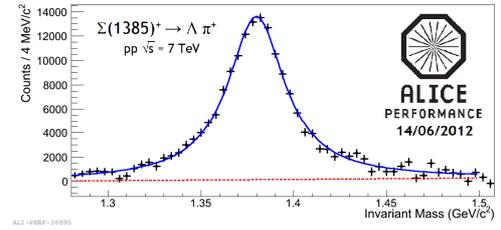

Figure 3: Σ$^{*}$ signal after the background subtraction. The fitting function (in blue) is the sum of a polynomial and a Breit-Wigner. The background polynomial from the fit is dashed in red.

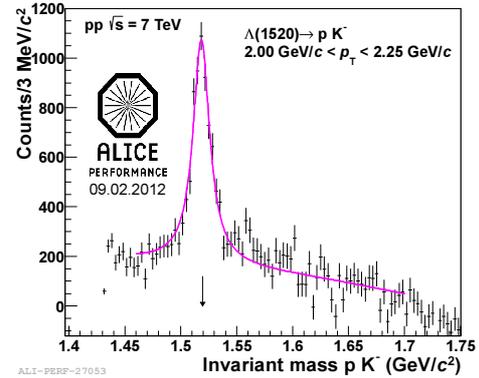

Figure 4: Λ$^{*}$ signal after the background subtraction. The fitting function (in magenta) is the sum of a polynomial and a Breit-Wigner.

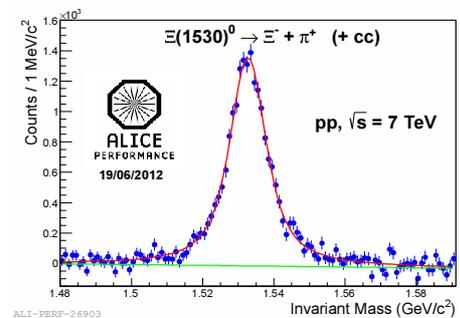

Figure 5: Ξ$^{*}$ signal after the background subtraction. The fitting function (in red) is the sum of a polynomial and a Voigtian. The background polynomial from the fit is dashed in green.

---

[1]The Voigtian is the convolution of a Gaussian and a Breit-Wigner and it is to account for the resolution in the invariant mass which ranges between 1 and 2 MeV/$c^2$ and is therefore comparable with the nominal φ(1020) and Ξ(1530) width (4.25 MeV and 9.1 MeV respectively).

[2]The results presented in this and in the following sections are the preliminary ones as shown during the conference. In the meantime the following preprint has been made available [13].

The spectrum is fitted with a Levy-Tsallis function [14]. The $\langle p_T \rangle$ is shown in Figure 7, as a function of the particle mass and compared with the value at 900 GeV from ALICE [15] and at 200 GeV from STAR [2, 4]. The $\langle p_T \rangle$ is also in agreement with the trend drawn by other particles at 7 TeV from ALICE [7, 8], which in turn differs from the ISR parametrization of the old values from lower energy experiments.

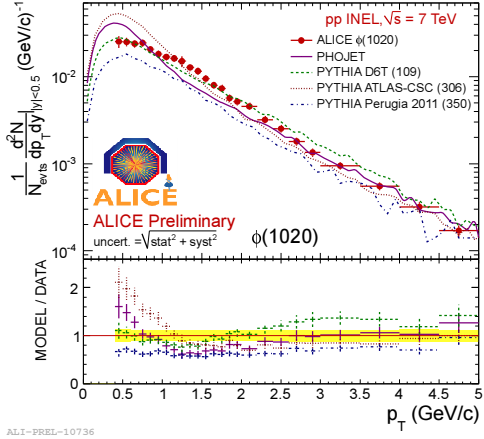

Figure 6: The transverse momentum spectrum of φ is compared to PHOJET and various PYTHIA tunes (D6T (109), ATLAS-CSC (306) and Perugia-2011 (350)).

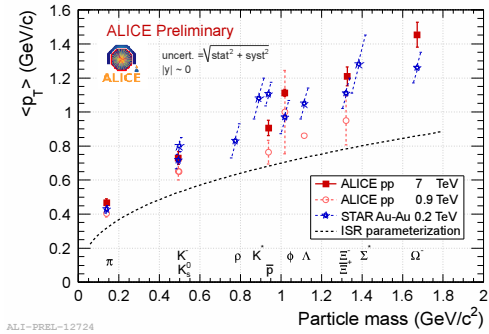

Figure 7: $\langle p_T \rangle$ vs. particle mass.

## 5. Comparison to the models

The transverse momentum spectra of φ and Σ* are compared to PHOJET and various PYTHIA tunes in Figures 6 and 8. For PYTHIA, tunes D6T (109), ATLAS-CSC (306) and Perugia-2011 (350) were used. For the φ the best agreement is found for the PYTHIA Perugia-2011 tune, which reproduces the high $p_T$ part ($p_T > 3$ GeV/$c$) of the φ spectrum rather well. PHOJET and ATLAS-CSC significantly overestimate the low momentum part ($p_T < 1$ GeV/$c$) of the transverse momentum distribution but reproduce the high momentum distribution. Finally, PYTHIA D6T tune is only able to describe the low-$p_T$ part ($p_T < 2$ GeV/$c$) of the φ spectrum. For the Σ* all the models seem to underestimate the data, with only a fair agreement for ATLAS-CSC at high $p_T$ ( > 2 GeV/$c$).

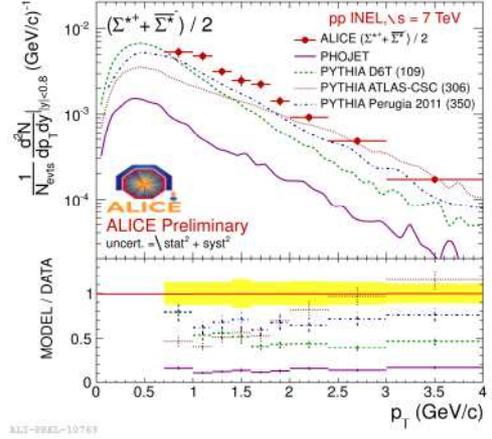

Figure 8: The transverse momentum spectrum of Σ* is compared to PHOJET and various PYTHIA tunes (D6T (109), ATLAS-CSC (306) and Perugia-2011 (350)).

## 6. Particle Ratios

Particle ratios in pp collisions are important as a baseline for comparison with heavy-ion reactions, where yields are in general strongly modified. In heavy-ion collisions, the yields for stable and long lived hadrons reflect the thermodynamic conditions (temperature, chemical potentials) during particle freeze-out, whereas the yields for short-lived, broad resonances like Σ* can be modified in addition by final state interactions inside the hot and dense reaction zone. These ratios are shown in Figures 9 and 10 together with the results obtained at lower incident energies in pp, $e^+e^-$, and A-A collisions. The φ/$K^-$ ratio is essentially independent of energy and also independent of the collision system. On the contrary, the φ/$\pi^-$ ratio increases with energy both in heavy-ion and in pp collisions up to at least 200 GeV.

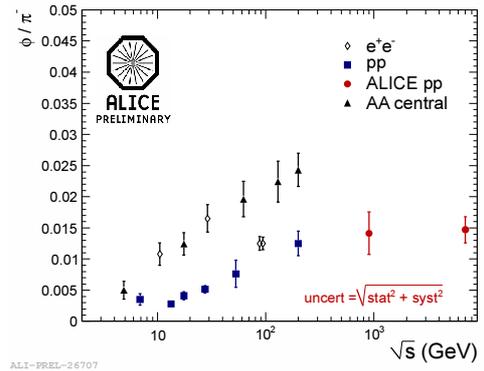

Figure 9: Energy dependence of the φ/$\pi^-$ in nuclear (triangles) [4, 17–21, 24] and pp (squares) [2, 4, 18, 21, 22, 25, 26] collisions. Diamonds represent the ALICE data at 0.9 and 7 TeV. The pion yields at 7 TeV are from [16]. The φ and $\pi^-$ yields at 0.9 TeV are from [23, 27].

In microscopic models where soft particle production is governed by string fragmentation, strange hadron yields are predicted to depend on the string tension [28]. Multi-strange baryons, and in particular the ratio Ω/φ, are expected to be very sensitive to this effect [29]. The φ yield is compared with the $\Omega^- + \Omega^+$ data measured by ALICE at the same colliding energy [9] in Figure 11 as a function of transverse momentum. The full line represents the PYTHIA model (Perugia 2011 tune), which is a factor 1.5-5 below the data. While this tune describes the φ spectrum reasonably well above 2-3 GeV/$c$, it under-predicts multistrange baryon yields by a large factor. The dashed line, which is very close to the data, represents the prediction of a model with increased string tension, the HIJING/BB v2.0 model with a Strong Colour Field [29] for pp collisions at 5.5 TeV. The value of string tension used in this calculation is K=2 GeV/fm, equal to the value used to fit the high baryon/meson ratio at $\sqrt{s}$ = 1.8 TeV reported by the CDF collaboration [30].

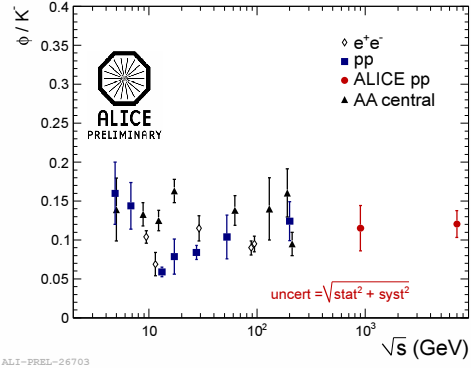

Figure 10: Energy dependence of the φ/K⁻ ratio in nuclear (triangles) [4, 17–21, 24] and pp (squares) [2, 4, 18, 21, 22, 25, 26] collisions. Diamonds represent the ALICE data at 0.9 and 7 TeV. The φ and K⁻ yields at 0.9 TeV are from [23, 27].

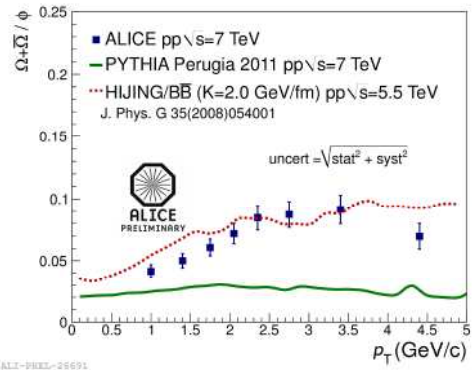

Figure 11: $(\Omega^- + \Omega^+)/\phi$ ratio as a function of transverse momentum for pp collisions at $\sqrt{s}$ = 7 TeV. The dashed line represents the prediction of HIJING/BB v2.0 model with a SCF for pp collisions at $\sqrt{s}$ = 5.5 TeV with a string tension of 2 GeV/fm [29]. The full line represents the prediction of the PYTHIA Perugia 2011 tune [31] for pp collisions at $\sqrt{s}$ = 7 TeV.

## 7. Conclusions

The hadronic resonances K*⁰(892), φ(1020), Σ(1385)±, Λ(1520), Ξ(1530) have been measured by the ALICE experiment in pp collisions at an energy in the center of mass of $\sqrt{s}$ = 7 TeV. Several methods for the raw yield extraction have been discussed and the values for the masses and the widths, when not constrained, were in agreement to the PDG values.

The transverse momentum spectra for the φ(1020) and Σ(1385) have been presented. A Levy-Tsallis function well describes the spectra. The yields of the φ(1020) have been compared to a previous measurement performed with ALICE at 900 GeV, showing an increase which scales with the multiplicity of the collisions. The average $p_T$ increases by about 30% with respect to previous measurements at 200 GeV. The φ/K⁻ ratio seems to stay independent of energy up to 7 TeV. Also the φ/π⁻ ratio, which increases in both pp and A-A collisions up to at least RHIC energies, seems to saturate and to become independent of energy above 0.2 TeV. The data have been compared to a number of PYTHIA tunes and the PHOJET event generator. None of them gives a fully satisfactory description of the data. The latest PYTHIA version (Perugia-2011) comes closest while still under predicting the φ meson $p_T$ spectrum below 3 GeV/$c$ by up to a factor of two. The $(\Omega^- + \Omega^+)/\phi$ ratio is not reproduced by PYTHIA Perugia 2011, but is in agreement with a prediction of HIJING/BB v2 model with a Strong Color Field with pp collision at 5.5 TeV which enhanced multistrange baryon production by increasing the string tension parameter.